\documentstyle[preprint,aps,graphicx]{revtex}

\def\red{
}
\def\black{
}

\def\URLtilde{\lower0.2em\hbox{$\tilde{\phantom{a}}$}}
\def\mycomm#1{\hfill\break\strut\kern-3em{\red\tt ====> #1\black}\hfill\break}
\def\mycommNL#1{\strut\kern-3em{\tt ====> #1}\hfill\break}
\def\ds{\displaystyle}

\catcode`\@=11 % This allows us to modify PLAIN macros.
\def\lsim{\mathrel{\mathpalette\@versim<}}
\def\gsim{\mathrel{\mathpalette\@versim>}}
\def\@versim#1#2{\vcenter{\offinterlineskip
           \ialign{$\m@th#1\hfil##\hfil$\crcr#2\crcr\sim\crcr } }}
\catcode`\@=12 % at signs are no longer letters

\def\sbar{\hbox{$\bar s$}}

\def\eqref#1{(\ref{#1})}

\makeatletter
\def\hlinewd#1{\noalign{\ifnum0=`}\fi
\hrule \@height #1 \futurelet \reserved@a\@xhline}
\def\hwhiteline{\noalign
{\ifnum0=`}\fi\hrule
\@height 0pt\vskip 1.0ex\futurelet \reserved@a\@xhline}
\makeatother
\def\gray{\special{ps: 0.40 setgray}}
\def\black{\special{ps: 0.0 setgray}}

\newcommand{\mydraft}{
\newcount\timecount
\newcount\hours \newcount\minutes  \newcount\temp \newcount\pmhours

\hours = \time
\divide\hours by 60
\temp = \hours
\multiply\temp by 60
\minutes = \time
\advance\minutes by -\temp
\def\hour{\the\hours}
\def\minute{\ifnum\minutes<10 0\the\minutes
       \else\the\minutes\fi}
\def\clock{
\ifnum\hours=0 12:\minute\ AM
\else\ifnum\hours<12 \hour:\minute\ AM
\else\ifnum\hours=12 12:\minute\ PM
       \else\ifnum\hours>12
        \pmhours=\hours
        \advance\pmhours by -12
        \the\pmhours:\minute\ PM
        \fi
       \fi
\fi
\fi
}
\def\fullclock{\hour:\minute}
\begin{centering}
\gray
\font\Hugett  =cmtt12 scaled\magstep4
\hbox{\Hugett Draft:\today,\clock}
\black
\end{centering}
\vskip -1.7cm
$\phantom{a}$
} % end of \draft definition
\def\beq#1{\begin{equation} \label{#1}}
\def\eeq{\end{equation}}
\def\bra#1{\left\langle #1\right\vert}
\def\ket#1{\left\vert #1\right\rangle}

\newskip\humongous \humongous=0pt plus 1000pt minus 1000pt

\newif\ifdtup

\def\TT{\hbox{\small $\bar{\bf 3}{\bf 3}$}}
\def\SS{\hbox{\small $\bar{\bf 6}{\bf 6}$}}
\def\3s{\hbox{\small $\bar{\bf s}\bar{\bf 3}$}}
\def\6s{\hbox{\small $\bar{\bf s}{\bf 6}$}}

\begin{document}
{\tighten
    \preprint {\vbox{
     \hbox{$\phantom{aaa}$}
     \vskip-0.5cm
%\hbox{\today}
%\hbox{}
\hbox{TAUP 2817-05}
\hbox{WIS/01/06-JAN-DPP}
\hbox{ANL-HEP-PR-06-4}
}}

\title{Diquarks and antiquarks in exotics: \\
a m\'enage \`a trois and a m\'enage \`a quatre}
\author{Marek Karliner\,$^{a}$\thanks{e-mail: \tt marek@proton.tau.ac.il}
\\
and
\\
Harry J. Lipkin\,$^{a,b}$\thanks{e-mail: \tt
ftlipkin@weizmann.ac.il} }
\address{ \vbox{\vskip 0.truecm}
$^a\;$School of Physics and Astronomy \\
Raymond and Beverly Sackler Faculty of Exact Sciences \\
Tel Aviv University, Tel Aviv, Israel\\
\vbox{\vskip 0.0truecm}
$^b\;$Department of Particle Physics \\
Weizmann Institute of Science, Rehovot 76100, Israel \\
and\\
High Energy Physics Division, Argonne National Laboratory \\
Argonne, IL 60439-4815, USA\\
}
\maketitle
%\mydraft

\begin{abstract}%
\strut\vskip-1.0cm

A m\'enage \`a trois is very different from an ordinary family. Similarly,
exotic hadrons with both  $qq$ and $q \bar q$ pairs have important color-space
correlations
that are
completely absent in ordinary mesons and baryons.
The presence of both types of
pairs requires attention to the basic QCD physics  that
the $q \bar q$ interaction is much stronger than the $qq$ interaction.
  This new physics in multiquark systems
produces  color structures  totally different from those of normal
hadrons, for example the $ud$ system is utterly unlike the $ud$
diquark in the $uds$ $\Lambda$ baryon. The color-space
correlations  produce unusual experimental properties in
tetraquarks with heavy quark pairs which may be relevant for newly
discovered mesons like the $X(3872)$ resonance. Tetraquark masses
can be below the two-meson threshold for sufficiently high quark
masses. A simple model calculation shows the  $b q \bar b \bar u$
and $b q \bar c \bar q$ tetraquarks below the $B \bar B$ and $B
\bar D$ thresholds. Some of these states have exotic electric
charge and their decays might have striking signatures involving
monoenergetic photons and/or pions.

\end{abstract}% %}
% end tighten

\vfill\eject

\section{Color, space and spin correlations in multiquark systems}

A reliable treatment of multiquark states should include the very different
physics arising in systems containing both quarks and antiquarks. The effects
of color-space correlations generally neglected was already pointed out in one
of the first multiquark treatments\cite{triex}. These can change the energy of
a state at least as much if not more than the color-flavor-spin correlations
commonly used\cite{NewPenta,JW,Jaffe}. A consistent treatment of confinement in
multiquark states must also consider that only color is confined and that 
color-singlet clusters are not confined. New additional states  having other
spin, color and space couplings not present in standard meson and baryon
spectroscopy must be included in realistic calculations. Such states may even
now be observed as exotic hadrons. Much of this essentially new  multiquark
physics is missed in many treatments. 

In three-quark baryons the color coupling of the three quarks to a color
singlet is unique and is antisymmetric in color. The Pauli principle then
requires the state to be totally symmetric in flavor, space and spin. An
isoscalar $ud$ state in a baryon; e.g. the $ud$ in the $\Lambda$ is required to
be a color antitriplet with spin zero.

The physics is completely different in states where both quark-quark and
quark-antiquark pairs are present. The color coupling is no longer unique. At
least two  color couplings are allowed and these can mix. No Pauli principle
connects quarks and antiquarks; all combinations of flavor, space and spin
symmetries can occur. Since the $q\bar q$ interaction in QCD is  much stronger
than the $qq$ interaction an antiquark between the two quarks in a quark-quark
system can completely change its color and space correlations; e.g. by mixing
in the color sextet state that does not exist in normal baryons.

This new physics is already evident in the color-space correlations absent in
normal hadrons appearing in the simplest multiquark system,  an isoscalar $ud
\bar s$ color triplet ``triquark" state. The two allowed color couplings have a
space-symmetric $ud$ state either in a color $\bar 3$ with spin zero, like the
diquark in a baryon, or a color $6$ with spin one, which
does not exist in three-quark baryons. These two couplings have been used in
calculations for the $uudd\sbar$ pentaquark where the $ud \bar s$ system must
couple to a color triplet which then couples to an additional isoscalar $ud$
diquark to make a color singlet. The pentaquark states with these two couplings
are nearly degenerate and therefore mix\cite{jenmalt}.

The triquark  state with the lowest potential energy was already shown to have
color couplings with the $qq$ in a color $6$, where the interaction is
repulsive, with this repulsion overwhelmed by a much stronger $q\bar q$
interaction\cite{triex,NewPenta}. The potential energy can then be lowered even
further by breaking maximum space symmetry and choosing a
configuration\cite{triex} where the mean distance between $qq$ pairs is much
larger than the mean distance between $q\bar q$ pairs. This is in contrast with
the 3q color singlet state in normal baryons where there is only a
single state. The use of diquarks in multiquark states with the same  color
couplings as the diquark in the baryon is therefore unjustified and leaves out
important physics.

One of the first treatments\cite{triex} of multiquark
systems considered a  tetraquark  system of two quarks and two
antiquarks with the Nambu color-exchange interaction\cite{Nambu}
now used to describe the spin-independent part of the interaction
in most common QCD-motivated
models\cite{NewPenta,JW,Jaffe,DGG,Lipflasy}.
The interaction between two constituents $i$ and $j$ is
\beq{Nambu}
    V_{cx}^{ij} = V \boldmath{\lambda}_c^i\cdot\lambda_c^j
\end{equation}
where $i$ and $j$ can be either quarks or antiquarks, $V$ is an
operator that depends on the space and spin variables of the
constituents but is the same for all pairs, independent of $i$ and
$j$, $\lambda_c^i$ is the generator of the color SU(3) group
and the scalar product $\lambda_c^i\cdot\lambda_c^j$ denotes the
scalar product in color space.

Nambu had already shown that no multiquark states are bound with this
interaction when spin effects are neglected, color and space are factorized
and all two-body subsystems have the same two-body density matrix. All color
singlet couplings of the same multiquark states have the  same potential
energy.  No potential energy is lost by coupling them into separate
noninteracting color-singlet meson and baryon clusters which then separate.

We generalize Nambu's theorem from color singlet states to any multiquark
system whose interaction has color dependence described by eq. (\ref{Nambu}).

\beq{Nambutot}
V_{cx}(tot) = \sum_{i \not=
j}{V_{cx}^{ij}\over2}= \sum_{i \not= j}
%\boldmath
{V\lambda_c^i\cdot\lambda_c^j\over2}
\end{equation}
For the case where the wave function factorizes into a color factor
and a factor depending on the other degrees of freedom the factor $V$
can be taken outside the summation to give
\beq{Nambutot2} V_{cx}(tot)
={V\over2} \cdot  \left[ \sum_{i j}
%\boldmath
\lambda_c^i\cdot\lambda_c^j -\sum_{i}
%\boldmath
(\lambda_c^i)^2\right]
={V\over2}
\cdot \left[\boldmath (\lambda_C)^2 -\sum_{i}\boldmath
(\lambda_c^i)^2\right]
\end{equation}
where
%\beq{casimir}
$\lambda_C =\sum_{i}\boldmath \lambda_c^i$
%\end{equation}
is the generator of the color SU(3) group for the whole multiquark system.

The interaction energy of a multiquark system depends only on the color of the
whole system and all states with the same overall color are degenerate. The
degeneracy can be removed by breaking factorization with color-spin and
color-space correlations. However, all the previously degenerate states remain
as physical states and  must be included.

The simple tetraquark model\cite{triex} broke the color-space factorization and
gained potential energy over the separated clusters by choosing a color
coupling where the $qq$ and $\bar q \bar q$ couplings were repulsive and
making the mean distance between  $qq$ and $\bar q \bar q$ larger than the mean
distance between $q \bar q$ pairs.  When the mean distances were equal the
enhanced $q \bar q$ attraction exactly overcompensated for the  $qq$ and  $\bar
q \bar q$ repulsion and gave this configuration the same potential energy as two
separated color singlet clusters. Making the distances unequal reduced the $qq$
and $\bar q \bar q$  repulsion relative to the $q \bar q$ attraction and
lowered the potential energy.

The present paper treats this tetraquark model in more detail and shows that it
can be relevant to presently observed mesons. A soluble toy model which
includes the kinetic energy and describes the spatial dependence of the
interaction (\ref{Nambutot}) by a harmonic oscillator potential exhibits the
effects of differences in spatial structure between exotics and normal hadrons.
These effects are often neglected or improperly treated in common treatments of
multiquark states which focus on color, flavor and spin and do not consider
spatial correlations.

\section{Color couplings in the \lowercase{$\boldmath(ud \bar s)$} system}

We first investigate the isoscalar color-triplet three-body $ud \bar s$ system. In the
simplest approximation with the Nambu two-body color-exchange interaction
(\ref{Nambu}-\ref{Nambutot}) and kinetic energies neglected, there are two
degenerate states and no unique definition of a state that can be called a
``diquark".

We consider the interactions in three possible color couplings: the two
orthogonal states denoted by  $\ket{\3s}$ and $\ket{\6s}$ which have an
isoscalar $ud$ pair coupled  either to a color $\bar {\bf 3}$ or
a color ${\bf 6}$ and
their ``quark-meson"  linear combination $\ket{d K^+}$ with the $u \bar s$
coupled to a color singlet $K^+$. We do not consider the $\ket{u K^o}$ state
as it has the same color physics as $\ket{d K^+}$. The state  $\ket{\6s}$ has a
$ud$ pair in a color sextet state which does not arise in ground state
baryons,
because it cannot couple with the third quark to a color singlet.  The $qq$
interaction in this state is repulsive; however, the attractive interaction of
the antiquark overcomes this repulsion in the three-body $(ud \bar s)$
system.

Eq. (\ref{Nambu}) relates the various two-body couplings to color singlet, antitriplet, sextet and octet states.
\beq{Namburel}
V_{cx}^{1} = 2V_{cx}^{\bar 3} = -4V_{cx}^{6} = -8V_{cx}^{8}
\end{equation}

The potential energy in a
potential model with the color-exchange two-body force (\ref{Nambu})
is given by
\beq{qbardiq}
\displaystyle
\bra {\3s}V\ket{\3s} =\phantom{-}
\displaystyle
 {1\over 2}\cdot\langle V_{ud}\rangle +
   {1\over 4}\cdot\langle V_{u\bar s} + V_{d\bar s}\rangle
= \langle V_{u\bar s}\rangle + {1\over 2}\cdot\langle V_{ud} -
V_{u\bar s}\rangle
\end{equation}

\beq{triq}
\displaystyle
\bra{\6s} V\ket{\6s}
\displaystyle
=
\displaystyle
-{1\over 4}\cdot\langle V_{ud}\rangle +
  {5\over 8}\cdot\langle V_{u\bar s}
+ V_{d\bar s}\rangle
= \langle V_{u\bar s} \rangle -{1\over 4}\cdot\langle V_{ud} - 
V_{u\bar s}\rangle
= {5\over 4}\langle V_{u\bar s} \rangle -{1\over 4}\cdot\langle V_{ud}\rangle
\end{equation}
\beq{dkaon}
\displaystyle
\bra{d K^+} V\ket{d K^+}
\displaystyle
 =
\displaystyle
 \langle V_{u\bar s}\rangle
\end{equation}
where $\langle V_{ud}\rangle$,
$\langle V_{u\bar s} \rangle$ and
$\langle V_{d\bar s}\rangle $
denote the expectation values of the three two-body potentials, the coefficients
are determined by the color algebra and we have used the isospin relation
\beq{isospin}
\langle V_{u\bar s} \rangle
= \langle V_{d\bar s}\rangle
\end{equation}

We see again the Nambu result that all couplings have the same color-electric
energy if the spatial wave functions are the same for all pairs; i.e. $\langle
V_{ud}\rangle = \langle V_{u\bar s}\rangle = \langle V_{d\bar s}\rangle$.
In this approximation,commonly used in most quark models for the
pentaquark, kinetic energies and color-space correlations are neglected and an
energy difference between different color couplings can only be due to the
color magnetic interaction.

There is no unique diquark-antiquark wave function for the three-body system.
The color sextet diquark, which does not exist in the three-quark color singlet
system, appears in the $qq \bar q$ system on the same footing as the color
antitriplet.

The $q\bar q$ interaction in the state $\ket{\6s}$ is seen to be 25\%
stronger than the $q\bar q$ interaction in the separated 
``quark-meson" state $\ket{d K^+}$. This additional attraction is balanced
exactly by the $ud$ repulsion if the spatial wave functions are the same for all
pairs. We thus see that additional attraction can be obtained by making the mean
$ud$ distance larger than the mean $q\bar q$ distance. We attempt to obtain a
quantitative estimate of this additional attraction by using a toy-model for the
spatial dependence of the interaction. 
 
We now go beyond this oversimplified treatment with a simple model which
includes kinetic energies and color-space-correlations. These drastically
modify the color states described by eqs. (\ref{qbardiq}) and (\ref{triq}).
These two states as well as the separated quark-kaon state (\ref{dkaon}) may no
longer be eigenstates of the new interaction. Our treatment does not include
this mixing and simply calculates the expectation values of the model
Hamiltonian including the color-space correlations for these three states with
the same color couplings. By the variational principle these give upper bounds
for the exact eigenvalues of the model Hamiltonian.

\section{Spatial and kinetic energy effects in the harmonic oscillator model}

We can estimate the spatial and kinetic energy effects by introducing a
spatial  dependence of the operator $V$  in the interaction (\ref{Nambutot}).
We choose the harmonic oscillator potential which enables easy separation of
center-of-mass motion and enables simple analytical exact solutions and
assume a nonrelativistic kinetic energy. Our model Hamiltonian is thus

\beq{VHO}
H = \sum_i {{p_i^2}\over{2m}} -
\sum_{i \not= j}
%\boldmath
{{V_o}\over{ 4}}\cdot\lambda_c^i\cdot\lambda_c^j \cdot r_{ij}^2 =
\sum_i {{p_i^2}\over{2m}} +{{V_o}\over{ 2}}\left[
\sum_{i j}
%\boldmath
\cdot\lambda_c^i\cdot\lambda_c^j \vec r_i \cdot \vec r_j -
\sum_i \lambda_c^i \cdot \sum_j\lambda_c^j r_j^2 \right]
\end{equation}
where $p_i$  denotes the momentum of particle $i$,  $\vec {r_{ij}}$ denotes
the distance between particles $i$ and $j$, and $m$  denotes the particle mass.
In this toy model we assume all masses to be equal.

This model Hamiltonian has been extensively investigated\cite{greenlip} and
been used in many multiquark calculations\cite{triex,jenmalt,hosaka}. Although
the harmonic oscillator potential becomes infinite at large distances, these
infinities are not relevant to our calculations which use harmonic oscillator 
Gaussian wave functions confined to regions where the potentials are
reasonable. 

We now introduce this explicit spatial form (\ref{VHO}) for the interactions into
eqs. (\ref{qbardiq}), (\ref{triq}) and (\ref{dkaon}) while keeping the same
color couplings. The potential for the separated $dK$ system becomes

%\beq{V3s}
%\begin{array}{ccl}
%\displaystyle
%{{V_{\3s}}\over{V_o}} = \displaystyle\phantom{-}
%{{1}\over{4}}\cdot r_{ud}^2 +
%{{1}\over{8}}\cdot r_{\bar s u}^2+ {{1}\over{8}}\cdot r_{\bar s d}^2
%\end{equation}
%$\hfill \\
%\\
%
%\beq{V6s}
%\displaystyle
%{{V_{\6s}}\over{V_o}} = \displaystyle -{{1}\over{8}}\cdot r_{ud}^2 +
%{{5}\over{16}}\cdot r_{\bar s  u}^2+ {{5}\over{16}}\cdot r_{\bar s d}^2
%\end{equation}
%
%\hfill \\
%\\
\begin{equation}
\displaystyle
{{V_{dK}}\over{V_o}} = \displaystyle\phantom{-}
{{1}\over{2}} \cdot (r_{\bar s u}^K)^2
%\end{array}
\end{equation}

For the triquark states we separate the center-of mass motion by
introducing the relative co-ordinates  

\begin{equation}
\begin{array}{ccl}
\displaystyle
\vec r &\equiv &\displaystyle \vec r_{ud}
=
 \vec r_u - \vec r_d; \qquad\qquad \vec R = \vec r_{\bar s} -
{{1}\over{2}}\cdot
(\vec r_u + \vec r_d) =
{{1}\over{2}}\cdot (\vec r_{\bar s u } + \vec r_{\bar s d})
%\end{equation}
\hfill\\
\\
%\begin{equation}
\displaystyle
{{V_{\3s}}\over{V_o}} &=&\displaystyle\phantom{-}
{{1}\over{4}}\cdot r^2 +
{{1}\over{8}}\cdot  [\vec R - (\vec r/2)]^2 +
[\vec R + (\vec r/2) ]^2 =
{{1}\over{4}}\cdot  R^2 + {{5}\over{16}}\cdot r^2
%\end{equation}
\hfill\\
\\\
%\begin{equation}
\displaystyle
{{V_{\6s}}\over{V_o}} &=&\displaystyle
-{1\over 8}\cdot r^2 +
{{5}\over{16}}\cdot  [\vec R - (\vec r/2)]^2 +
[\vec R + (\vec r/2) ]^2=
{{5}\over{8}}\cdot   R^2  + {1\over 32}\cdot r^2
\end{array} 
\end{equation}

The potentials are seen to be positive definite functions of the variables $R$
and $r$ and therefore bounded from below. 
Although the color sextet has a repulsive force, the negative term $-(1/8)\cdot
r^2$ is overcompensated by the attraction of the antiquark and the resulting
dependence upon $r^2$ has a positive coefficient. 
That the Hamiltonian (\ref{VHO}) is positive definite for overall color singlet
and triplet states was pointed out in general for a
color-exchange interaction with a harmonic oscillator force\cite{greenlip}. This
is easily seen by inspection from eq. (\ref{VHO}) for an overall color singlet
since $\sum_{i j}
\lambda_c^i\cdot\lambda_c^j \vec r_i \cdot \vec r_j $ is a square and 
$\sum_i \lambda_c^i =0$ for a color singlet.
 
The expectation values of the model hamiltonian  (\ref{VHO})
give the energies of the ground states for these three systems
\begin{equation}
\begin{array}{ccccc}
\displaystyle
E_{dK} &=&
\bra{d K^+} H \ket{d K^+} &=&
\displaystyle
{{3\hbar}\over{2}}\cdot
    \omega^{dK};
\hfill \\
\\
\displaystyle
E_{\bar s \bar 3}&=&
\bra {\3s} H \ket{\3s} &=&
 1.40  \cdot E_{dK};
\hfill \\
\\
\displaystyle
E_{\bar s 6}&=& \bra{\6s} H \ket{\6s} &=& 1.22\cdot  E_{dK}
\end{array}
\end{equation}
where $\omega^{dK}$ is the oscillator frequency for the separated quark-kaon system
\begin{equation}
\omega^{dK} =
\sqrt{2V_o\over m_{\hbox{\scriptsize\rm reduced}}} =
\sqrt{2V_o\over m}
\end{equation}

The ground state mean square distances are
\begin{equation}
\label{rsquareddK}
\begin{array}{c}
\displaystyle
\strut\kern6em
\langle [r^{dK}]^2 \rangle ={3\hbar\over m\omega^{dK}}=
{{3\hbar}\over {\sqrt
{mVo}}} \cdot{{\sqrt 2}\over {2}}
%
%\end{equation}
\hfill\\
\hfill\\
\\
%\begin{equation}
\displaystyle
\langle [r^{\3s}]^2 \rangle =
1.26\cdot \langle [r^{dK}]^2 \rangle
; \qquad\qquad
\langle [r_{\bar su}^{\3s}]^2 \rangle =
1.54 \cdot\langle [r^{dK}]^2 \rangle
%\label{rsquaredqdiq}
\hfill\\
\hfill\\
\\
%\end{equation}
%\begin{equation}
\displaystyle
\langle [r^{\6s}]^2 \rangle =
4 \cdot \langle [r^{dK}]^2 \rangle
; \qquad\qquad
\strut\kern1em
\langle [r_{\bar su}^{\6s}]^2 \rangle =
1.77 \cdot\langle [r^{dK}]^2\rangle
%\label{rsquaredtriq}
\\
\strut
\end{array}
\end{equation}

 The results in eq.~(\ref{rsquareddK}) are
schematically illustrated
in Fig.\,1 below.
\vbox{
\strut\vskip0.5cm
\strut\kern6em
\includegraphics[width=11.0cm,angle=90]{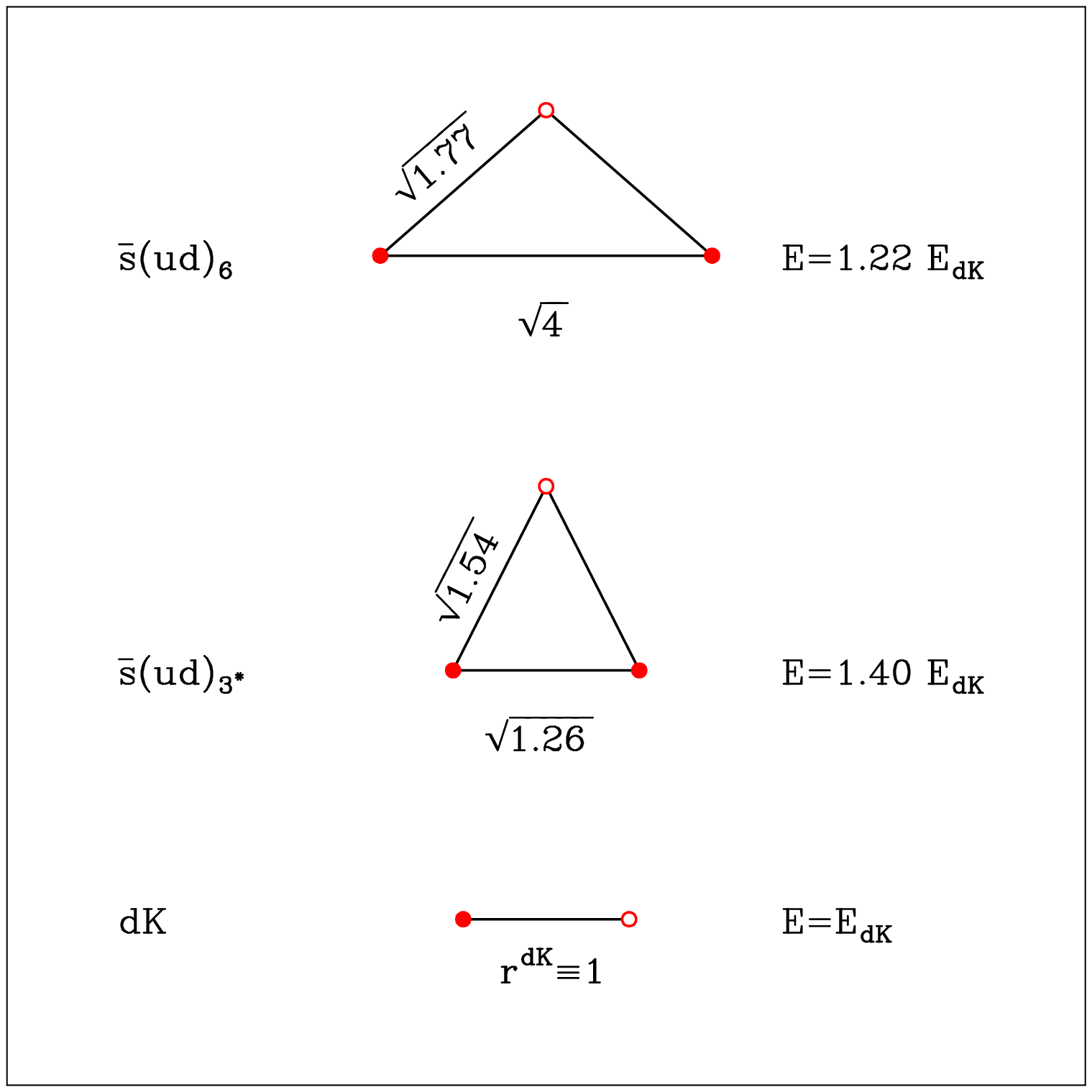}
\hfill\break
\hfill\break
{\small\it
\baselineskip=0pt
Fig. 1. Schematic depiction of the geometry of the $dK$, $\bar
s(ud)_{3^*}$ and $\bar s(ud)_{6}$ configurations. Quarks are denoted by full
circles, antiquarks by empty circles. The
lengths of the triangles' sides are square roots of the corresponding
rms distances. In the $dK$ configuration the ``outside" $d$-quark is not shown,
as there are no color forces between it and the color-singlet $K$ meson.}

\hfill\break
\strut\vskip0.5cm
}

The $\ket{\6s}$ triquark  has a
lower energy than the $\ket{\3s}$ triquark  but still
higher than the energy of the separated quark-kaon system. The mean square
quark-quark and quark-antiquark distances in both triquarks  are larger than
the quark-antiquark distance in the kaon. They are largest for the $\ket{\6s}$
triquark.

When color-space factorization is abandoned, the various color  recouplings
produce different energies and rms distances.  The interaction of the two
quarks with the antiquark creates a $ud$ structure very different from that
of the $ud$ diquark found in three-quark baryons.

These results for the $(u d \bar s)$ system are not easily tested against
experiments which always involve color singlet hadrons. We now extend  the
basic physics of this approach to treat realistic exotic multiquark color
singlet systems; e.g. tetraquarks and pentaquarks, for which there are
experimental candidates and for which multiquark models have already been
proposed.

The case of the $(u u d d \bar s)$ pentaquark has been treated \cite{hosaka}
using a similar harmonic oscillator hamiltonian with an additional
spin-dependent interaction. This system is much more complicated than our
triquark because it includes two pairs of quarks with identical flavor and the
necessity to properly antisymmetrize the wave function. They are therefore
unable to find a simple set of relative co-ordinates in which the spatial
problem can be solved exactly and center-of-mass motion can be ignored. They
use a model space of single-particle shell-model wave functions in a model
space of 15,000 basis  functions.

One solution found in ref. \cite{hosaka} has a color-space
correlation similar to that of our triquark. The mean
quark-antiquark distance is much smaller than the mean quark-quark
distance. In particular they find that the one-body r.m.s. radius
measured from the center of mass is 1.10 Fm. for $u$,$d$ quarks
and 0.72 Fm for the $\bar s$ antiquark while the corresponding
radius is 0.69 Fm for the $(0s)^5$ configuration.

That the one-body r.m.s. radius of the multiquark system is larger than
expected from normal hadrons implies that the short-range color-magnetic
interaction will be weaker than in normal hadrons. The conventional practice
of using spin splittings from normal hadrons to normalize the color-magnetic
interaction\cite{NewPenta,JW,Jaffe} is therefore questionable.

\section{A Tetraquark in the harmonic oscillator model}

We consider the simplest multiquark color singlet system which can be
treated exactly in the harmonic oscillator model without color mixing and
without the need for Pauli antisymmetrization. The tetraquarks treated here
exhibit features missed in other models which have been suggested for these
states; e.g. the color sextet quark-quark couplings and can provide useful
models for mesons containing heavy quarks that are now being discovered.

As discussed in more detail in Sec. V,
these wave functions may also provide a good
basis for further treatments in which color-magnetic and spin effects are
treated as perturbations.

\subsection{The $(u s \bar d \bar c)$ tetraquark}

We develop general results for tetraquarks by considering the $(u s \bar d \bar
c)$ system with four different flavors and four different masses. This will
show how the difference between the tetraquark mass and the  mass of two
separated mesons depends upon the quark masses.

The general multiquark model hamiltonian (\ref{VHO}) used for the interaction
used in the three triquark states  (\ref{qbardiq}), (\ref{triq}) and
(\ref{dkaon}) is now used for the analogous four-body states. There is a
separated two-meson state denoted by $2M$. 

\begin{equation}
\label{pot2M}
{{V_{2M}}\over{V_o}} = {{1}\over{2}} \cdot (r_{u \bar c}^2 + r_{s\bar d})^2
\end{equation}

There are a triplet-antitriplet tetraquark
denoted by $\TT$, and a sextet-antisextet
tetraquark denoted by $\SS$. 

\begin{equation}
\label{pottetra}
\begin{array}{cc}
\ds
{{V_{\TT}}\over{V_o}} =
\phantom{-}
{{1}\over{4}}\cdot (r_{us}^2 +
r_{\bar c \bar d}^2)+ {{1}\over{8}}\cdot (r_{u \bar c}^2+ r_{s \bar c}^2 +
r_{u \bar d}^2+ r_{s \bar d}^2)
\hfill\\
\\
%\end{equation}
%\begin{equation}
\ds
{{V_{\SS}}\over{V_o}} = -{{1}\over{8}}\cdot (r_{us}^2 +
r_{\bar c \bar d}^2)+ {{5}\over{16}}\cdot (r_{u \bar c}^2+ r_{s \bar c}^2 +
r_{u \bar d}^2+ r_{s \bar d}^2)
\end{array}
%\hfill\\
%\\
\end{equation}

Here again we note that when all the spatial separations are equal,
$r_{us}^2 = r_{\bar c \bar d}^2 = r_{u \bar c}^2= r_{s \bar c}^2 =
r_{u \bar d}^2= r_{s \bar d}^2$
the $q\bar q$ interaction in the state $\SS$ is seen to be 25\%
stronger than the $q\bar q$ interaction in the separated 
two-meson state $2M$. This additional attraction is balanced
exactly by the $us$ and $\bar c \bar d$ repulsions if the spatial wave 
functions are the same for all
pairs. We thus see that additional attraction can be obtained by making the mean
$us$ and $\bar c \bar d$ distances larger than the mean $q\bar q$ distance. 

We also note that the repulsive $qq$ interaction in the color-sextet plays a
crucial role in the suppression of binding low-lying exotic hadrons. The
additional $q\bar q$ attraction arises from the presence of four attractive
$q\bar q$ interactions in the tetraquark, whereas there are only two attractive
$q\bar q$ interactions in the separated two-meson system. The larger number of
$q\bar q$ pairs in multiquark systems as opposed to separated color singlet
hadrons would produce bound low-lying exotics if there were no repulsive $qq$
interactions.

We now obtain a 
quantitative estimate of this additional attraction by using the harmonic
oscillator model for the spatial dependence.

We express the interactions $V_{\TT}$ and $V_{\SS}$ in terms of
the relative co-ordinates
%\beq{rel1}
\begin{equation}
\label{potDD}
\begin{array}{c}
\ds
\strut\kern4em
\vec r_{us} = \vec r_u - \vec r_s; ~ ~ ~ \vec r_{dc} = \vec r_{\bar d} - \vec
r_{\bar c} ; ~ ~ ~ \vec R ={{1}\over{2}}\cdot (\vec r_{\bar c} + \vec r_{\bar
d}) - {{1}\over{2}}\cdot (\vec r_u + \vec r_s)
%\end{equation}
\hfill\\
\\
%\beq{potDD}
\ds
{{V_{\TT}}\over{V_o}}
=
\phantom{-}
{{1}\over{4}}\cdot (r_{us}^2 +
    r_{cd}^2)+{{1}\over{2}}\cdot R^2 +{{1}\over{8}}\cdot (r_{us}^2 +
r_{cd}^2 ) = {{1}\over{2}}\cdot R^2 + {{3}\over{8}}\cdot (r_{us}^2 +
r_{cd}^2 )
\hfill\\
\\
%\end{equation}
%\beq{pottetra}
\ds
{{V_{\SS}}\over{V_o}} = -{{1}\over{8}}\cdot (r_{us}^2 +
    r_{cd}^2)+{{5}\over{4}}\cdot R^2 +{{5}\over{16}}\cdot (r_{us}^2 +
r_{cd}^2 ) = {{5}\over{4}}\cdot R^2 + {{3}\over{16}}\cdot (r_{us}^2 +
r_{cd}^2 )
\end{array}
\end{equation}

The ground state energies of these states are given by the sum of the ground
state energies of the contributing harmonic oscillators. The ground state
energy of an oscillator with co-ordinate $r$, mass $M$
and potential $\,Vr^2\,\,$  is

    \beq{gs}
E_g(osc) = {{3}\over{2}}\cdot\hbar \omega =
{{3}\over{2}}\cdot\hbar\sqrt{{{V}\over{M}}}
\end{equation}

For the two-meson
$(u \bar c;s \bar d$)
system, which has two separated harmonic oscillators with reduced masses
denoted by $M(uc)$ and
$M(ds)$ and potentials
$V_o/2$ from eq.(\ref{pot2M}),
    \beq{E2Msucd} E_g(2Musdc) = {{3}\over{2}}\cdot\hbar
\left(\sqrt{{{V_o}\over{2M(uc)}}}+
\sqrt{{{V_o}\over{2M(ds)}}}\,\,\right)
\end{equation}

The ground state energies for the $\TT$ and $\SS$ systems are obtained by
substituting the potentials in (\ref{potDD}) into the
expression (\ref{gs}) along with  the reduced masses

    \beq{redmassc}
M(us)={{m_u m_s}\over{m_u +m_s}}; ~ ~ ~  M(dc)={{m_d m_c}\over{m_d +m_c}};
    ~ ~ ~ M\{R;(us)(dc)\}
    = {{(m_u +m_s)\cdot(m_d +m_c) }\over{m_u +m_s + m_d +m_c}}
\end{equation}

\beq{EDD}
E_g(\TT) = {{3}\over{2}}\cdot\hbar
\left(\sqrt{{{V_o}\over{2M\{R;(us)(dc)\}}}}+
\sqrt{{{3V_o}\over{8M(us)}}}+
\sqrt{{{3V_o}\over{8M(dc)}}}\,\,\,\right)
\end{equation}
\beq{Etetra1}
E_g(\SS) = {{3}\over{2}}\cdot\hbar
\left(\sqrt{{{5V_o}\over{4M\{R;(us)(dc)\}}}}+
\sqrt{{{3V_o}\over{16 M(us)}}}+
\sqrt{{{3V_o}\over{16 M(dc)}}}\,\,\,\right)
\end{equation}
The ratios of these energies to the energy of the two meson state are then

\beq{EDDsucdrat0}
\begin{array}{l}
\displaystyle
{{E_g(\TT\,usdc)}\over{E_g(2M\,usdc)}} =
%\displaystyle
%\sqrt{3\over4}+
%\left[{{\sqrt{M(dc)M(us)}}\over{
%$\sqrt{M(us) M\{R;(us)(dc)\}}+
%\sqrt{M(dc) M\{R;(us)(dc)\}}}}\,\,\right]
\\
\hfill\\
%\end{equation}
%\beq{EDDsucdrat3}
%{{E_g(\TTusdc)}\over{E_g(2M\,usdc)}}
%\displaystyle
%=
%\displaystyle
%\sqrt{3\over4}
%+{{\sqrt{(m_u +m_s + m_d +m_c)m_sm_um_cm_d}}\over{
%\sqrt{(m_u +m_s)\cdot (m_d +m_c)}
%\cdot
%\displaystyle
%\left[
%\sqrt{m_u m_s\cdot (m_d +m_c)}+
%\sqrt{m_d m_c\cdot (m_u +m_s)}\,\,\right]} }
%\\
%\hfill\\
%\end{equation}
%\beq{EDDsucdrat3}
%{{E_g(\TTusdc)}\over{E_g(2M\,usdc)}}
%\displaystyle
%=
\displaystyle
\sqrt{3\over4}
+{{\sqrt{(m_u +m_s + m_d +m_c)m_sm_um_cm_d}}\over{m_cm_s(m_d-m_u)
 + m_dm_u (m_c-m_s)}}
\cdot
\displaystyle
\left[\sqrt{{{m_d m_c}\over{m_d +m_c}}}-
\sqrt{{{m_u m_s}\over{m_u +m_s}}}
\,\,\,\right]
=1.18
\end{array}
\end{equation}

\beq{Etetra(cusurat)}
\begin{array}{c}
\displaystyle
{{E_g(\SS\,usdc)}\over{E_g(2Musdc)}} =
%\sqrt{{{3}\over{8}}}+
%\sqrt{5\over 2}\cdot
%\left[{{\sqrt{M(dc)M(us)}}\over{
%\sqrt{M(us) M\{R;(us)(dc)\}}+
%\sqrt{M(dc) M\{R;(us)(dc)\}}}}\right]
%\end{equation}
\hfill\\
\\
%\beq{Etetracusurat2}
\displaystyle
\strut\kern-1em
= \sqrt{{{3}\over{8}}}+
\sqrt{5\over 2}\cdot
{{\sqrt{(m_u +m_s + m_d +m_c)m_sm_um_cm_d}}\over{m_cm_s(m_d-m_u)
 + m_dm_u (m_c-m_s)}}
\cdot
\displaystyle
\left[\sqrt{{{m_d m_c}\over{m_d +m_c}}}-
\sqrt{{{m_u m_s}\over{m_u +m_s}}}
\,\,\,\right]
=1.11
\end{array}
\end{equation}
where we have set $m_u = m_d$ and substituted the  values of
the constituent quark masses obtained by
fitting the ground state meson and baryon spectra\cite{NewPenta}.

\beq{massval} m_u =  360 ; ~ ~ ~
m_s=  540  ; ~ ~ ~ m_c= 1710 ; ~ ~ ~  m_b =  5050
\end{equation}

The ratios of the energy of each tetraquark state to the energy of the
separated two-meson system are the sum of two terms.
The first terms, respectively $\sqrt {3/4}$ and $\sqrt {3/8}$ are the energies
of the $qq$ and
$\bar q \bar q$ systems which respectively have the same reduced masses as the
corresponding mesons. They have a lower energy because they are spread over a
larger domain in configuration space and have a lower kinetic energy. The
second terms depend upon the flavor via the reduced mass of the
$(qq)-(\bar q \bar q)$ system. These ratios decrease with increasing quark
masses and for sufficiently high quark masses will
produce bound tetraquarks below the two-meson threshold, as discussed
in Secs. C and D below.

\subsection{The $c u \bar c \bar u$ tetraquark}

We treat the $c u \bar c \bar u$ configuration
by changing the flavors in eqs. (\ref{EDDsucdrat0})--(\ref{Etetra(cusurat)}).
This gives

\beq{EDDcucurat}
{{E_g(\TT\,cucu)}\over{E_g(2M\,cucu)}}
%= \sqrt{{{3}\over{4}}}+\sqrt{{{M(cu)}\over{4M\{R;(cu)(cu)\}}}}
%= \sqrt{{{3}\over{4}}}+{{\sqrt{\medstrut 2m_c m_u}}\over{\medstrut 2(m_c +m_u)}}
=1.134 ; ~ ~ ~
%\end{equation}
%\beq{Etetra(cucurat)}
{{E_g(\SS\,cucu)}\over{E_g(2Mcucu)}} 
%\sqrt{{{3}\over{8}}}+\sqrt{{{5M(cu)}\over{8M\{R;(cu)(cu)\}}}}
%= \sqrt{{{3}\over{8}}}+{{\sqrt{\medstrut 5m_c m_u}}\over{2(m_c +m_u)}}
=1.036
\end{equation}
where we have substituted the constituent quark masses (\ref{massval}).

That the energy of the $c u \bar c \bar u$ tetraquark with the color
sextet-antisextet coupling $(\SS)$ is found to be only 4\% above the ground
state energy of the two-meson $D\bar D$ system suggests that such states should
be included in all calculations for these states.

That the $(\SS)$ state is considerably below the $(\TT)$ state casts doubt on
all tetraquark calculations which neglect the color space
correlations\cite{Hogaasen} and the $(\SS)$ state\cite{diqdiqbar,eric}; e.g.
those for the X(3872) resonance.   These neglect the basic physics seen in the
experimental hadron mass spectrum and its description by QCD motivated
models\cite{NewPenta,JW,Jaffe,DGG,Lipflasy} showing that the attractive  $q\bar
q$ interaction as observed in mesons is much stronger than the attractive $qq$
interaction observed in baryons.

The color-space correlation contributions to the energy may well be more
important than  the color-magnetic energy neglected here which dominates other
tetraquark model calculations\cite{diqdiqbar,Hogaasen,eric}.

The color magnetic energy
can be included in the same way as in all other quark model calculations, as a
perturbation using the unperturbed wave functions. The correction may not be
large here because of the small color-magnetic interaction of the heavy charmed
quark and the suppression of this  short-range interaction at larger distances.
This is discussed in detail in section V.

\subsection{The $b u \bar b \bar u$ tetraquark}

We treat the $b u \bar b \bar u$ configuration
by changing the flavors in eqs. (\ref{EDDsucdrat0})--(\ref{Etetra(cusurat)}).
This gives
\beq{EDDbuburat}
{{E_g(\TT\,bubu)}\over{E_g(2M\,bubu)}}
%= \sqrt{{{3}\over{4}}}+\sqrt{{{M(bu)}\over{4M\{R;(bu)(bu)\}}}}
%= \sqrt{{{3}\over{4}}}+{{\sqrt{\strut 2m_b m_u}}\over{2(m_b +m_u)}}
=1.042; ~ ~ ~
%\end{equation}
%\beq{Etetra(buburat)}
{{E_g(\SS\,bubu)}\over{E_g(2M\,bubu)}} 
%\sqrt{{{3}\over{8}}}+\sqrt{{{5M(bu)}\over{8M\{R;(bu)(bu)\}}}}
%= \sqrt{{{3}\over{8}}}+{{\sqrt{\strut 5m_b m_u}}\over{2(m_b +m_u)}}
= 0.891
\end{equation}
where we have substituted the constituent quark masses (\ref{massval}).

The mass of the $b u \bar b \bar u$ tetraquark with the color sextet-antisextet
coupling is found to be well below the two-meson $B\bar B$ threshold in this
approximation. Such a low-mass threshold may also be found in a more exact
calculation including spin effects. That they might be found in the experimental
spectrum must be taken seriously.

\subsection{The $c u \bar b \bar u$ tetraquark}

We treat the $c u \bar b \bar u$ configuration
by changing the flavors in eqs. (\ref{EDDsucdrat0})--(\ref{Etetra(cusurat)}).
This gives
\beq{EDDcuburat}
%\begin{array}{c}
\displaystyle
{{E_g(\TT\,cubu)}\over{E_g(2M\,cubu)}} =
%\beq{EDDsucdrat3}
%{{E_g(\TTusdc)}\over{E_g(2M\,usdc)}}
%\displaystyle
%=
%\displaystyle
%\sqrt{3\over4}
%+{{\sqrt{(m_b + 2m_u +m_c)m_bm_c}}\over{m_u (m_b-m_c)}}
%\cdot
%\displaystyle
%\left[\sqrt{{{m_um_b}\over{m_u +m_b}}}-
%\sqrt{{{m_um_c}\over{m_u +m_c}}}
%\right]
=1.10; ~ ~ ~
%\\
%\hfill\\
%\end{array}
%\end{equation}
%
%\beq{Etetra(cuburat)}
%\begin{array}{c}
\displaystyle
{{E_g(\SS\,cubu)}\over{E_g(2M\,cubu)}} 
%\sqrt{{{3}\over{8}}}+
%\sqrt{5\over 2}\cdot
%\left[{{\sqrt{M(bu)M(cu)}}\over{
%\sqrt{M(cu) M\{R;(bu)(cu)\}}+
%\sqrt{M(bu) M\{R;(bu)(cu)\}}}}\right]
%\\
%\hfill\\
%\end{equation}
%\beq{EDDsucdrat3}
%{{E_g(\TTusdc)}\over{E_g(2M\,usdc)}}
%\displaystyle
%=
%\displaystyle
%\sqrt{{{3}\over{8}}}+
%\sqrt{5\over 2}\cdot
%{{\sqrt{(m_b + 2m_u +m_c)m_bm_c}}\over{m_u (m_b-m_c)}}
%\cdot
%\displaystyle
%\left[\sqrt{{{m_um_b}\over{m_u +m_b}}}-
%\sqrt{{{m_um_c}\over{m_u +m_c}}}
%\right]
= 0.975
%\end{array}
\end{equation}
where we have substituted the constituent quark masses (\ref{massval}).

This suggests that a
$c u \bar b \bar u$ tetraquark might exist with a mass at the $BD$
threshold. This could be a narrow state decaying electromagnetically to
the $B_c$ and a photon or an electron pair. A spin-zero tetraquark with a
mass too low for pion decay to $B_c$ would have to decay by electron pair
emission in a $0\rightarrow 0$ transition. It is therefore of interest to
search for such tetraquarks in experiments observing the $B_c$.

%   \beq{DDbar}
%E_g(D\bar D) = {{1}\over{2}}\cdot2\hbar \left(\sqrt{{{9
%V_o}\over{2M(uc)}}} =
%{{1}\over{2}}\cdot2\hbar \sqrt{{{23 V_o}\over{1520}}}\right)
%\end{equation}

%\beq{EDD(c u c u}
%E_g(DDcucu) =
%{{1}\over{2}}\cdot\hbar\left(\sqrt{{{9V_o}\over{M\{R;(uc)(uc)\}}}}
%+\sqrt{{{3\cdot 9 V_o}\over{2M(cu)}}}\right)=
%{{1}\over{2}}\cdot\hbar\left(\sqrt{{{ V_o}\over{230}}}
%+\sqrt{{{69V_o}\over{1520}}}\right)
%\end{equation}
%\beq{Etetracucu}
%E_g(\SScucu) = {{1}\over{2}}\cdot\hbar\left(\sqrt{{{9\cdot
%5V_o}\over{4M\{R;(uc)(uc)\}}}}+
%\sqrt{{{9\cdot 3V_o}\over{4 M(cu)}}}\right)=
%{{1}\over{2}}\cdot\hbar\left(\sqrt{{{5V_o}\<<<<<<<<<<<<<<<<<<<<<<<<<<<<<<over{460}}}+
%\sqrt{{{69V_o}\over{3040}}}\right)
%\end{equation}

\subsection{The $s u \bar b \bar u$ tetraquark}

We treat the $s u \bar b \bar u$ configuration
by changing the flavors in eqs. (\ref{EDDsucdrat0} - \ref{Etetra(cusurat)})
and using the constituent quark masses (\ref{massval}) as before.
This gives
\beq{EDDsuburat}
% \begin{array}{c}
\displaystyle
{{E_g(\TT\,subu)}\over{E_g(2M\,subu)}} =
%  \sqrt{3\over4}+
%  \left[{{\sqrt{M(bu)M(su)}}\over{
%  \sqrt{M(su) M\{R;(bu)(su)\}}+
%  \sqrt{M(bu) M\{R;(bu)(su)\}}}}\right]
%  \\
%  \hfill\\
%  \displaystyle
%  =
%  \displaystyle
%  \sqrt{3\over4}
%  +{{\sqrt{(m_b + 2m_u +m_s)m_bm_s}}\over{m_u (m_b-m_s)}}
%  \cdot
%  \displaystyle
%  \left[\sqrt{{{m_um_b}\over{m_u +m_b}}}-
%  \sqrt{{{m_um_s}\over{m_u +m_s}}}
%  \right] =
1.16 ; ~ ~ ~
%  \end{array}
%\end{equation}
%
%\beq{Etetra(suburat)}
%  \begin{array}{c}
\displaystyle
{{E_g(\SS\,subu)}\over{E_g(2M\,subu)}} =
%  \sqrt{{{3}\over{8}}}+
%  \sqrt{5\over 2}\cdot
%  \left[{{\sqrt{M(bu)M(su)}}\over{
%  \sqrt{M(su) M\{R;(bu)(su)\}}+
%  \sqrt{M(bu) M\{R;(bu)(su)\}}}}\right]
%  \\
%  \hfill\\
%  \displaystyle
%  =
%  \displaystyle
%  = \sqrt{{{3}\over{8}}}+
%  \sqrt{5\over 2}\cdot
%  {{\sqrt{(m_b + 2m_u +m_s)m_bm_s}}\over{m_u (m_b-m_s)}}
%  \cdot
%  \displaystyle
%  \left[\sqrt{{{m_um_b}\over{m_u +m_b}}}-
%  \sqrt{{{m_um_s}\over{m_u +m_s}}}
%  \right]=
1.077
%  \end{array}
\end{equation}
% where we have substituted the constituent quark masses (\ref{massval}).

\section{Inclusion of color magnetic interactions}

\subsection{Spin quantum numbers of tetraquarks}

The spin states of the $qq\bar q\bar q$ system where all pairs are in relative
$S$-waves can be
denoted as $\ket{s_q,s_{\bar q};S}$ where $s_q$,$ s_{\bar q}$,
and $S$ denote the spins respectively of the quarks, the antiquarks and the
total  spin.  The states of the
$S$-wave
four-body system are thus
$$
\hbox{
$\ket{1,1;0}$,
\qquad
$\ket{1,1;2}$
\qquad
and
\qquad
$\ket{0,0;0}$
}
$$
which can decay into two pseudoscalar mesons, and
$$
\hbox{
$\ket{1,1;1}$,
\qquad
$\ket{0,1;1}$
\qquad
and
\qquad
$\ket{1,0;1}$
}$$
whose decay into two pseudoscalars is forbidden by angular
momentum and parity.

We now estimate the corrections from color-magnetic and spin effects
to the above results in
the conventional manner used in constituent quark calculations\cite{Jaffe,DGG}.
The expression for the interaction energy (\ref{Nambu}) is generalised to
include an explicit spin dependence\cite{Jaffe,DGG} which is not the same
for all pairs. The interaction energy between two constituents
$i$ and $j$ is
now given by\cite{Jaffe}

\beq{Nambumag}
V_{cx}^{ij} = (V - {{16 v}\over {3}}\boldmath\vec \sigma_i \cdot \vec
\sigma_j)
  \cdot (\lambda_c^i\cdot\lambda_c^j)
\end{equation}
where $\vec \sigma$ denotes
the Pauli spin operators and $v$ is a strength parameter fixed by experiment.

We first note that when this interaction is treated as a first-order
perturbation, the color magnetic contribution calculated from
eq.(\ref{Nambumag})vanishes when suitably averaged over masses of the
pseudoscalar ${\cal P}$
and vector ${\cal V}$
 states~\cite{NewPenta,nudatlamb} of a quark of flavor $i$ bound to a
$\bar u$ antiquark.
\beq{nuX}
\ket {{\cal P}_i} = \ket {q_i \bar u}_{S=0}; ~ ~ ~ ~ ~
\ket {{\cal V}_i} = \ket {q_i \bar u}_{S=1}; ~ ~ ~ ~ ~
%\end{equation}
%\beq{nuY}
\tilde M({\cal V}_i) = {{3 M({\cal V}_i)+ M({\cal P}_i) } \over 4}
\end{equation}
where $\tilde M$ denotes the linear combination of masses for which the
color magnetic interaction (\ref{Nambumag})cancels out \cite{NewPenta,nudatlamb}.

We can immediately relate the calculated two-meson masses like
(\ref{E2Msucd})
to the experimental masses
    \beq{E2Mxcubu}
E_g(2Mcubu)=\tilde M(c\bar u) +\tilde M(b\bar u)= {{3 M(D^*)+ M(D) } \over
4}
+{{3 M(B^*)+ M(B) } \over 4}
\end{equation}
\beq{E2Mxcubu2}
E_g(2Mcubu)=M(D) + M(B) +{{3 [M(D^*)- M(D)] } \over 4}
+{{3 [M(B^*)- M(B)] } \over 4}
\end{equation}

Thus the statements about the $BD$ threshold must be corrected to include the
hyperfine splittings.
For
the $BB$ system,
 \beq{E2Mxbubu}
E_g(2Mbubu)=2 \tilde M(b\bar u)+ {{3 M(B^*)+ M(B) } \over 2}
=2 M(B) +{{3 [M(B^*)- M(B)] } \over 2}
\end{equation}

The $B^*-B$ mass difference is 46 MeV, the color-magnetic
mass reduction is 69 MeV for the $BB$ system and 23 MeV for the $B^*B$ system.
These are down in the noise of our rough calculation.

The color-magnetic corrections for the tetraquark masses are not easily
calculated because they depend upon the spatial wave functions, but the
sign of the change vs. ordinary hadrons is clear. A short-range
two-body hyperfine interaction depends upon the wave function at the origin,
which can be calculated using the harmonic oscillator wave functions. Since the
mean square distances are considerably larger in the tetraquarks  than in
mesons, a simple scaling of the wave function at the origin will scale the
hyperfine  splittings in the tetraquark to a lower value than those in the
observed mesons. The neglect of the hyperfine splittings is therefore justified
as a reasonable approximation, as supported by the following estimates.

\subsection{Some estimates of the color magnetic interaction}

 The lowest tetraquark state when color-magnetic interactions are
included is expected to be the spin zero state of a spin-one color sextet and a
spin-one antisextet, $\ket{1,1;0}$. Although the color-magnetic energy is not
easily calculated in the general case,
the quark-antiquark contribution for this state
is simplified because of its symmetry.

In the state $\ket{1,1;0}$ the spin of one of the antiquarks
must couple with the total spin 1 of the quarks to give a total spin 1/2. The
relevant angular momentum algebra gives the expectation values  of   $\vec
\sigma_q \cdot \vec\sigma_{\bar q}$ for a single $q\bar q$ pair in this state
and for any spin singlet state of the $q\bar q$ system
\beq{spinqqbar6}
\bra{1,1;0}\vec \sigma_q \cdot \vec\sigma_{\bar q} \ket{1,1;0}  = -2; ~ ~ ~ ~ ~
%\end{equation}
%\beq{spinsing}
\bra{(q\bar q)_{S=0}} \vec \sigma_q \cdot \vec\sigma_{\bar q} \ket{(q\bar
q)_{S=0}} = -3
\end{equation}

The contribution of the $q\bar q$ pairs to the color magnetic energy of this
state is obtained by substituting the spin expectation values (\ref{spinqqbar6})
into eq. (\ref{Nambumag}), summing over all quarks and antiquarks and
substituting the $SU(3)$ Casimir operators for the sextet
$C_3(6) = (40/3)$ and triplet $C_3(3) = (16/3)$.
\beq{colormag6}
\bra{1,1;0}V_{CM}(q\bar q) \ket{1,1;0} =
{{16 v}\over {3}}\cdot 2\cdot \bra{1,1;0}
\boldmath{\sum_{i=q}\sum_{j=\bar q}}
\lambda_c^i\cdot\lambda_c^j
 \ket{1,1;0} = {{16 v}\over {3}}\cdot 2\cdot C_3(6)
\end{equation}

The ratio of the color-magnetic energy for this state to the
color-magnetic energy of two mesons can be obtained for the case
where the spatial distances between quark-antiquark pairs is equal
in all cases,
 \beq{colormag6rat}
 {{\bra{1,1;0}V_{CM}(q\bar q) \ket{1,1;0}}\over {\bra{2M}V_{CM}(q\bar q) \ket{2M}}}
= {{ C_3(6)}\over {3\cdot C_3(3)}}= {{5}\over{12}}
\end{equation}

This suggests that the color-magnetic effects reduce the mass of the tetraquark
by an amount roughly half of the reduction in mass of the two-meson system.

\section{Experimental implications of the existence of tetraquarks}

\subsection{Tetraquark symmetry quantum numbers and decay selection rules}

States which have two quark-antiquark pairs of the same flavor have charge
conjugation quantum numbers,
and states having a light quark-antiquark pair
have isospin and $G$-parity quantum numbers. $SU(3)$ flavor symmetry is not
considered here as this symmetry is badly broken and considering
the effects of
symmetry-breaking on selection rules requires further analysis.

The $c u \bar c \bar u$ and $b u \bar b \bar u$ tetraquarks which
contain charge-conjugate sextet-antisextet states have charge
conjugation quantum numbers that are conserved in strong and
electromagnetic decays. Starting with the spin quantum numbers, we
note that the $\ket{s_q,s_{\bar q};S}$ states
 $\ket{1,1;0}$, $\ket{1,1;2}$ and
$\ket{0,0;0}$  are all even under $C$ and therefore also even
under $CP$. The $\ket{1,1;1}$ state is odd under $C$ and therefore
also odd under $CP$, while the $\ket{0,1;1}$ and $\ket{1,0;1}$ are
linear combinations of even and odd $C$ and the sum and difference
of these states are respectively even and odd under both $C$ and
$CP$.

The states having a light quark-antiquark pair are found with both isospins, 0
and 1. Since the mesons with one heavy or strange quark and one light quark have
isospin 1/2, two-meson states with isospin 0 and 1 exist and no isospin
selection rule forbids a tetraquark decay into two mesons. Charmonium and
bottomonium states have isospin zero. Thus the $\pi \Upsilon$ and $\pi J\psi$
strong decays are allowed only for tetraquarks with isospin one and are isospin
forbidden for isospin-zero tetraquarks.

The $c u \bar c \bar u$ and $b u \bar b \bar u$ tetraquarks which
contain a heavy quark-antiquark pair of the same flavor can decay
into light quark mesons by annihilation of the heavy
quark-antiquark pair.  Although a single $Q\bar Q$ state can decay
strongly into light quarks only by annihilating the $Q\bar Q$ into
two or three gluons, a $Q\bar Q$ constituent in a $Q \bar Q d\bar
u$ tetraquark can annihilate into one gluon which is then absorbed
in the light quark system. Such states are apt to be broad. These
include all tetraquarks with these constituents, including the
molecular states.

States like the $c u \bar b \bar u$ and $s u \bar b \bar u$ tetraquarks which
heavy quark-antiquark pairs of different flavors cannot decay strongly or
electromagnetically into light quark mesons. Their strong decays must conserve
their heavy flavors and isospin.  Thus the strong $\pi B_s$ and $\pi B_c$ decays are
are allowed only for tetraquarks with isospin one and are isospin
forbidden for isospin-zero tetraquarks.

\subsection{Experimental detection via decay modes}

Thus the best candidates for experimental detection are the $c u
\bar b \bar u$ and $s u \bar b \bar u$ tetraquarks which cannot
decay strongly or electromagnetically into light quark mesons.

     A $b q \bar c \bar q$ tetraquark with isospin 1 and a mass below the $B
\bar D$ threshold but above the mass of the $B_c \pi$ system can
decay strongly into a $B_c \pi$ and should have a strong width
limited by phase space. The $I=0$ state and the $I=1$ state below
the $B_c \pi$ threshold should be narrow. They can both decay
electromagnetically into a $B_c$ and a photon and the $I=0$ state
above the $B_c \pi$ threshold can  decay  into $B_c \pi$ via
isospin violation.  The above considerations apply also to states
below the $B^* \bar D$ threshold that cannot decay into $B\bar D$.

If the mass of a $b q\bar c \bar q$ tetraquark is below the $B_c$
mass it can decay only weakly. The tetraquark with the same charge
as the $B_c$ can have the same final state as the $B_c$; e.g.
$J/\psi e\nu$. It would appear in any invariant mass plot of the
final state as an additional mass peak along with the $B_c$.
Isovector tetraquarks like $b u \bar c \bar d$ or $b d \bar c \bar
u$ have exotic final states with wrong charges, like $J/\psi \eta$
or $J/\psi \pi^- \pi^-$.

There is therefore interest in looking for monoenergetic photons or pions
emitted together with a $B_c$ meson, a doublet structure of the $B_c$ mass and
exotic $B_c$ decays, like $J/\psi \eta$ or $J/\psi \pi^{\pm} \pi^{\pm}$.

    Analogous considerations hold for a $b q \bar s \bar q$ tetraquark with a
mass below the $B K$ threshold: strong decay and strong width into
$B_s \pi$ for $I=1$ states above $B_s \pi$ threshold; narrow
widths for all other states into $B_s \gamma$ or $B_s \pi$ with
isospin violation; weak decays for tetraquarks below the $B_s$
mass, producing a mass doublet with the $B_s$ or exotic final
states with wrong charges, like  $J/\psi \pi^{\pm}$.

A $c q \bar c \bar q$ tetraquark with a mass below the $D \bar D$
threshold or
 a $b q \bar b \bar q$ tetraquark with a
 mass below the $B \bar B$ threshold
would very likely decay strongly into light quark hadrons and be very broad.
The $I=1$ states might also decay strongly by pion emission
respectively
into $\pi J/\psi$ or $\pi \Upsilon$.
Isospin violating pion emission might occur for $I=0$ states above the pion
threshold.
Electromagnetic decays into  $\gamma J\psi$ or $\gamma \Upsilon$
might also occur.

    Since exotic multiquark states are required by color-space
correlations to have a larger extension in space than normal hadrons,
they may be easily broken  up by final state interactions or rescattering
by other particles in the same final state. This may make it difficult for
such states to be produced and survive in multiparticle final states.

\section{conclusion}

Color-space correlations can play a crucial role in multiquark systems and may
be more important than the color-flavor-spin correlations generally dominating
other treatments of multiquark states.

In any theoretical treatment of multiquark states containing both quarks and
antiquarks the full implications of the basic QCD physics that the $q\bar q$
interaction observed in mesons is much stronger than the $q q$ interaction
observed in baryons must be considered.  Admixtures of quark states that do not
exist in normal baryons\cite{NewPenta,jenmalt} must be included. In the simplest
multiquark system containing one antiquark and a quark system the $q\bar q$
interaction already destroys completely the spatial and color structures of the
quark system that existed in the absence of the antiquark including all diquark
structures. Color-space correlated tetraquarks may already be observed in
mesons containing heavy quarks.

The most likely signals that would indicate the unambiguous
presence of tetraquarks should be searched in experiments focused
on the $B_c$ system. The $b q \bar c \bar q$ tetraquark might be
below the $B\bar D$ threshold. The possible exotic signatures
include strong or electromagnetic decays into a $B_c$ and a pion
or photon, weak decays producing additional peaks in the mass
spectrum of $B_c$ decay final states or weak decays into states
with exotic electric charge, like $J/\psi \eta$ or $J/\psi \pi^{-}
\pi^{-}$.

\section*{Acknowledgements}

The research of M.K. was supported in part by a grant from the
Israel Science Foundation administered by the Israel
Academy of Sciences and Humanities.

%----------------------------------------------------------------------
% This prevents REFERENCES from forcing a page break
%\def\newpage{\vskip10ex}
%
\catcode`\@=11 % This allows us to modify PLAIN macros
\def\references{
\ifpreprintsty \vskip 10ex
%\ifpreprintsty \newpage
%
\hbox to\hsize{\hss \large \refname \hss }\else
\vskip 24pt \hrule width\hsize \relax \vskip 1.6cm \fi \list
{\@biblabel {\arabic {enumiv}}}
{\labelwidth \WidestRefLabelThusFar \labelsep 4pt \leftmargin \labelwidth
\advance \leftmargin \labelsep \ifdim \baselinestretch pt>1 pt
\parsep 4pt\relax \else \parsep 0pt\relax \fi \itemsep \parsep \usecounter
{enumiv}\let \p@enumiv \@empty \def \theenumiv {\arabic {enumiv}}}
\let \newblock \relax \sloppy
    \clubpenalty 4000\widowpenalty 4000 \sfcode `\.=1000\relax \ifpreprintsty
\else \small \fi}
\catcode`\@=12 % at signs are no longer letters
%-----------------------------------------------------------------
%{\tighten

} % end of global \tighten
\end{document}